\documentclass[journal]{IEEEtran}
\usepackage{graphicx,amssymb,lineno}
\usepackage{amsmath,amsfonts,amssymb}

\usepackage{algorithm}
\usepackage{algorithmic}
\usepackage[usenames]{color}
\usepackage{float}
\usepackage{booktabs}

\usepackage{graphicx,graphics,color,epsfig,subfigure,graphpap,rotate}
\usepackage{times, verbatim, subfigure, epsfig, graphicx, latexsym, amsmath}
\usepackage{url}
\usepackage{subfigure}
\begin{document}
\title{ Coalition Game based Full-duplex Concurrent Scheduling in Millimeter Wave Wireless Backhaul Network }
\author{Haiyan~Jiang, Yong~Niu, Jiayi~Zhang, Bo~Ai, and Zhangdui~Zhong
\thanks{H. Jiang, Y. Niu, J. Zhang, B. Ai, and Z. Zhong are with the State Key Laboratory of Rail Traffic Control and Safety, and the School of Electronic and Information Engineering, Beijing Jiaotong University, Beijing 100044, China (Corresponding author: Yong Niu, e-mails:niuy11@163.com).}
}
\maketitle
\begin{abstract}
With the development of self-interference (SI) cancelation technology, full-duplex (FD) communication becomes possible.
FD communication can theoretically double the spectral efficiency.  When the time slot (TS) resources are limited and the number of flows is large, the scheduling mechanism of the flows becomes more important. Therefore, the effectiveness of FD scheduling mechanism for the flows is studied in millimeter wave wireless backhaul network with the limited TS resources. We proposed a full duplex concurrent scheduling algorithm based on coalition game (FDCG) to maximize the number of flows with their QoS requirements satisfied. We transformed the problem of maximizing the number of flows with their QoS requirements satisfied into the problem of maximizing sum rate of concurrently scheduled flows in each slot. We obtained the scheduled flows with maximum sum rate in first slot by using coalition game.And then with certain restrictions, the maximum sum rate of concurrently scheduled flows can also be achieved in subsequent time slots. The simulation results show that the proposed FDCG algorithm can achieve superior performance in terms of the number of flows that meet their QoS requirements and system throughput compared with other three algorithms.
\end{abstract}


\section{Introduction}\label{S1}
In recent years, mobile data traffic has increased exponentially, and some industry and academic experts predict a 1000-fold demand increase by 2020 \cite{first}. To meet this requirement, a ultra-dense cellular network with small cells densely deployed become a promising solution for explosive traffic growth in the fifth-generation (5G) network \cite{second}. With a generous number of base stations (BSs) deployed, the massive backhaul traffic turned into a great challenge. Because wired optical fiber based backhaul network with large bandwidth is time-consuming, costly and inflexible in densely deployed small cells, wireless backhaul network will be widely used in 5G. The traditional microwave band (such as the frequency band below 5GHz) is limited to the achievable gain due to the existing spectrum tension problems. Although there are many new technologies that can improve the spectrum efficiency, it is still difficult to achieve the rate of over 1Gbps or even 10Gbps. Millimeter-wave (mmWave) bands has huge bandwidth and can reach multi-gigabit rates, so mmWave wireless backhaul can provide a more cost-effective and feasible solution for densely deployed small cells backhaul \cite{NY1}.

In order to compensate for high path loss in mmWave communication, directional antennas and beamforming techniques are often used \cite{Training}. The directional communication is difficult for the third party nodes to perform the carrier sense and solve the deafness problem \cite{NY3}. On the other hand, it reduces the interference among flows, so that concurrent transmission can be used to improve the network capacity to a great extent \cite{seventeen}. However, concurrent transmission of multiple flows lead to higher multi-user interference (MUI), which conversely affect the network performance. Therefore, how to effectively schedule the flows transmitted concurrently has drawn great attention of many researchers. For example, a dual update concurrent scheduling algorithm in rate-adaptive wireless networks was proposed in \cite{third}, but the influence of millimeter-wave characteristics was not taken into account. Then, in \cite{seven}, considering the characteristics of mmWave and the QoS requirements of flows, Qiao \emph{et al.} proposed a concurrent scheduling algorithm, which was called STDMA. And this algorithm selected the set of concurrent scheduling flows in a greedy way. But both non-interference flows and interference flows can be scheduled at the same time to maximize the throughput of the network, which reduce the rate of concurrent scheduling. Qiao \emph{et al.} \cite{six} proposed a time-slot resource sharing scheme for mmWave 5G cellular networks, in which D2D communication technology and concurrent scheduling technology were used to improve the network capacity. However, in order to simplify the problem, only non-interference flows can share resources in each slot. And this scheme doesn't consider the QoS requirements of flows. In order to improve the performance of the system, Niu \emph{et al.} proposed a joint transmission scheduling scheme of access and backhaul in mmWave small cell, called D2DMAC, to schedule the flows with the least time slots to meet the QoS requirements of flows \cite{five}. Although latency is reduced, a flow that needs more time slots will not be transmitted in subsequent slots. In \cite{eight}, Zhu \emph{et al.} introduced the concept of QoS-aware independent set on the scheduling problem in mmWave backhaul network, and proposed a method that can further improve the system throughput and increase the number of successfully transmition flows (that is, the number of flows with their QoS requirements satisfied) compared with STDMA, called MQIS (Maximum QoS-aware Independent Set). It doesn¡¯t remove the flow(s) spent too much slots, nor does it evaluate the profits when adding a new flow. Niu \emph{et al.} \cite{ten} studied the energy efficient scheduling for mmWave backhauling of small cells in heterogeneous cellular networks. The maximum independent set of each pair was obtained by minimum greedy algorithm, and then the number of slots to each pair and the transmission power to each flow were assigned by power control algorithm, so that energy efficient could be realized. In this paper, we remove the flow(s) spend too much slots and consider the QoS requirement of flow. We don't consider power control. Maybe our future research will involve power control in scheduling. The objective of this paper is to maximize the number of completed flows, i.e., the number of flows with the QoS requirements satisfied.

These concurrent scheduling schemes [7]-[12] in mmWave bands hold the assumption of half-duplex (HD). In recent years, with the development of self-interference (SI) cancellation technology \cite{eleven} \cite{twelve}, it becomes possible to enable full-duplex (FD) communication in mmWave wireless backhaul networks. But SI cannot be completely eliminated, there is still residual self-interference (RSI). Here, the SI means that the transmitted signal is received by the local receiver at the same base station (BS). It seriously affects the performance of the FD system \cite{eleven}. FD communication is achieved by transmitting and receiving information simultaneously at the same BS over the same
frequency, which may theoretically double the spectral efficiency and provides an important opportunity for concurrent scheduling in mmWave wireless backhaul networks  \cite{eleven} \cite{twelve}. In \cite{thirteen}, Feng \emph{et al.} introduced FD communication into the 5G mmWave backhaul network scheduling scheme, but the scheduling scheme was designed in the case of time slots enough. Therefore, for mmWave backhaul networks, how to satisfy the QoS requirements of flows as much as possible is still a challenge under the condition that the TS resources are limited. This is also the content of this paper to study. In \cite{Twety}, Ding \emph{et al.} proposed a QoS-aware FD concurrent scheduling scheme for mmWave wireless backhaul network with limited time slot resources. This scheme largely improves system throughout,but the number of flows that satisfy the QoS requirements compared with the total number of flow increase to a lesser extent. We consider the introduction of relative interference which determined whether flows can be scheduled concurrently causes some flows not to be scheduled, which reduces the number of flows. We choose to use the maximum sum rate to determine whether flows can be scheduled concurrently. When the sum rate of each slot reaches the maximum, the number of completed flows will reach the maximum.

Coalition game has also been widely explored in different disciplines such as economics or political science. Recently, cooperation has emerged as a new networking paradigm that has a dramatic effect of improving the performance of communication networks \cite{fourteen}. Coalition games prove to be a very powerful tool for designing fair, robust, practical and efficient cooperation strategies in communication networks \cite{fourteen}. In this paper, the coalition game is introduced to schedule the concurrent scheduling flows in mmWave wireless backhaul network.

The above opportunities and challenges motivate us to investigate a coalition game based full-duplex concurrent scheduling in millimeter wave wireless backhaul network. The contribution of this paper are summarized as follows.
\begin{itemize}

\item We formulate the optimization problem of concurrent scheduling in FD mmWave wireless backhaul network into a nonlinear inter programming (NLIP) problem, i.e., maximizing the number of flows with their QoS requirements satisfied in case of the limited time slot resources. Both RSI and MUI are simultaneously taken into account.There are scheduled concurrently flows with the maximum sum rate in each slot to achieve the completed flows as many as possible.

\item We propose a FD concurrent scheduling scheme based on coalition game (FDCG) in mmWave wireless backhaul network, which consists of a maximum independent set (MIS) algorithm that can be achieved by the minimum-degree greedy algorithm and a coalition game algorithm, to efficiently solve the formulated problem. FDCG uses the coalition game algorithm to select the set of concurrently scheduled flows with the maximum sum rate in the first slot. In order to guarantee the set of concurrently scheduled flows with the maximum sum rate in the subsequent slot, this scheme evaluate the profits when adding a new flow, i.e., when the sum rate of the set of flows after added a new flow is greater than that of the set of flows without added a new flow, this new flow will be added to scheduled concurrently.

\item Through evaluation under various system parameters, we demonstrate that our scheme can significantly increase the number of flows with their QoS requirements satisfied and improve the total system throughput. Furthermore, we also analyze respectively the impact of SI cancellation level, MUI factor and different QoS requirement on the performance improvement.

\end{itemize}

The structure of this paper is organized as follows. Section \ref{sec:System-Model} introduces system models and explain
the formulated optimization problem in detail. The coalition game based scheduling algorithm is presented in
Section \ref{sec:coalition game algorithm}. Section \ref{sec:result analysis} shows the simulation results and evaluates
 the performance compared with existing schemes. Finally, we make a conclusion about this paper in Section \ref{sec:conclution}.

\section{\label{sec:System-Model}SYSTEM MODEL AND PROBLEM FORMULATION}
\subsection{System Model}
\begin{figure}
\begin{center}
\includegraphics[width=8cm]{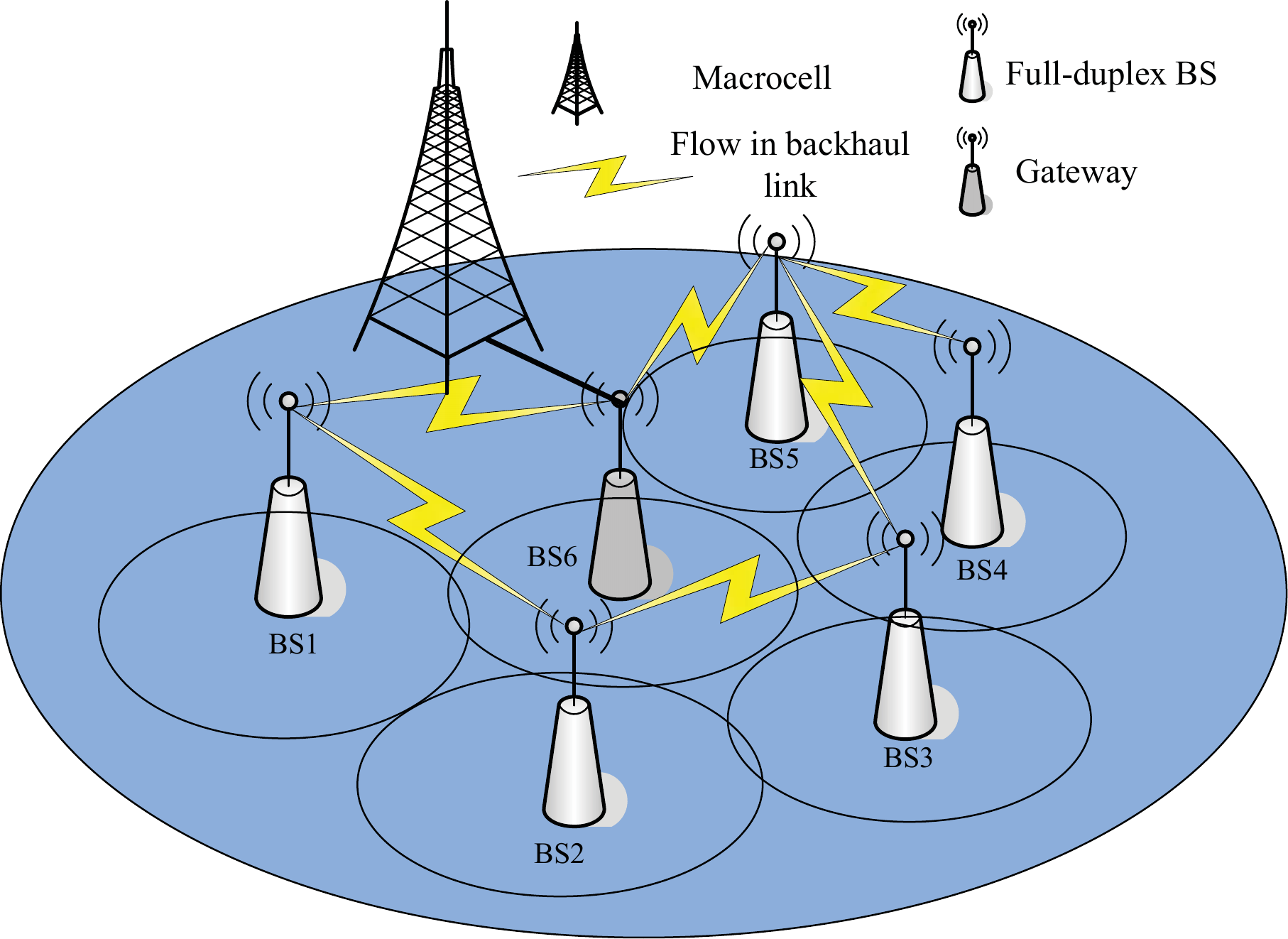}
\end{center}
\caption{Full-deplex mmWave wireless backhaul network in the small cells densely deployed scenario.}
\label{environment}
\end{figure}

\begin{figure}
\begin{center}
\includegraphics[width=8cm]{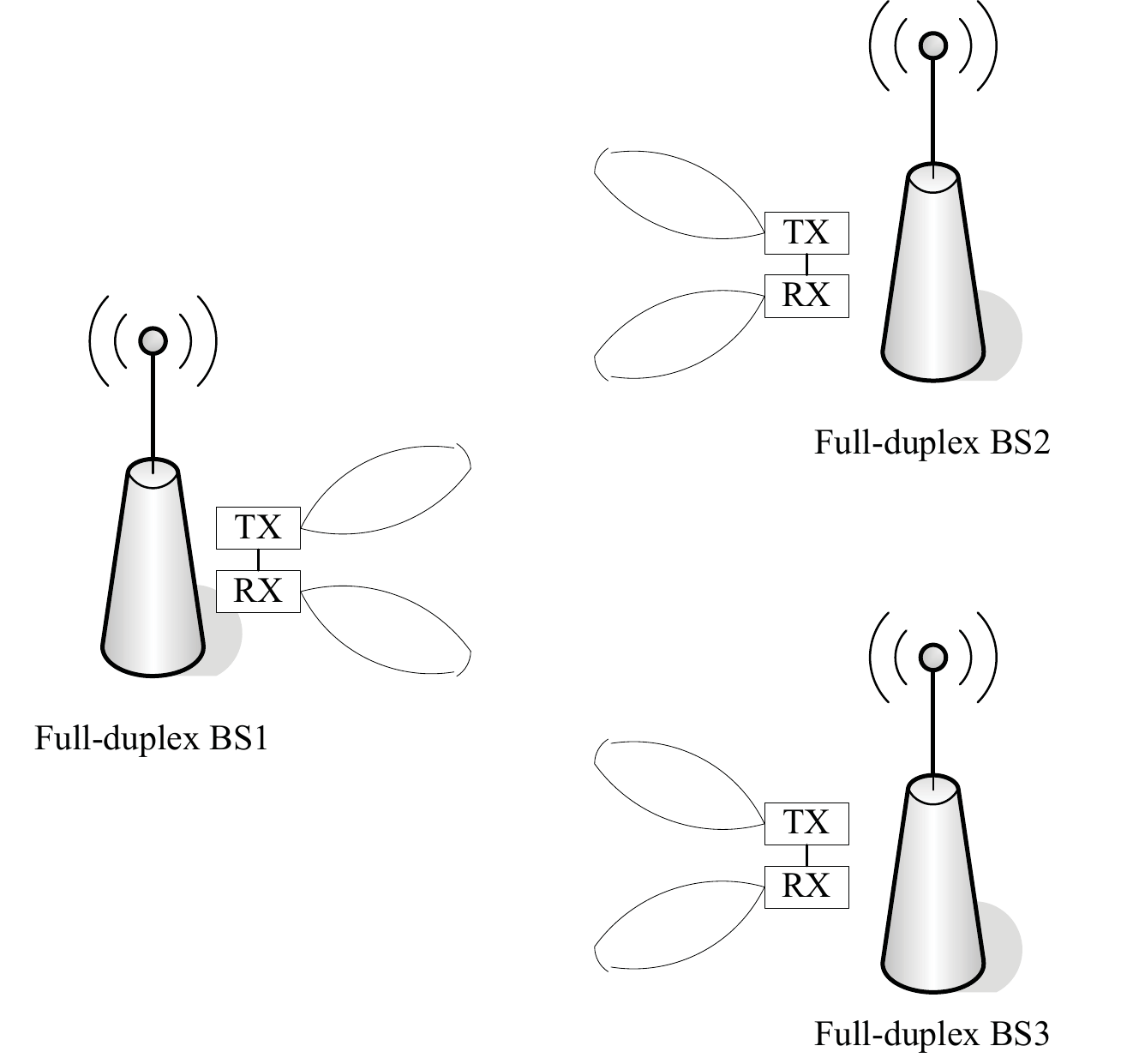}
\end{center}
\caption{Full-duplex base stations.}
\label{bs}
\end{figure}

We consider a typical FD mmWave wireless backhaul network in the small cells densely deployed scenario \cite{Twety}.
As shown in Fig. \ref{environment}, the network includes \emph{N} BSs (only six BSs are drawn in Fig.1) that are connected to each other by backhaul links with mmWave band. When there are some traffic demands from one BS to another, we say there is a flow between them. There are one or more BSs connected to the backbone network via the macrocell, which is (are) called gateway(s) \cite{ten}. A backhaul network controller (BNC) belongs to one of the gateways, which can synchronize the network, receive the QoS requirements of flows and obtain the locations of BSs \cite{fourteen}. In Fig. 1, BS6 is both a gateway and a full-duplex BS. Each BS operates in FD mode and is equipped with two steerable directional antennas: one for transmitting and another for receiving, as shown in Fig. \ref{bs}. Therefore, a BS can at most support two flows at the same time. It can simultaneously serve as the transmitter of one flow and the receiver of another, but it can¡¯t simultaneously serve as the transmitters or receivers of both two flows. When BS3 and BS4 send data to BS5 at the same time in Fig. 1, two flows will fail. If one flow is scheduled in next slot, another flow can be scheduled successfully in this slot. The effective scheduling scheme can make full use of  time slot resources and improve system throughput.

We adopt the widely used realistic directional antenna model, which is a main lobe of Gaussian from in linear scale and constant level of sidelobes \cite{fifteen}. It is the reference antenna model with sidelobe for IEEE 802.15.3c \cite{3cxie}. The gain of a directional antenna in units of decibels (dB), which is denoted by $G(\theta)$, can be expressed as
\begin{eqnarray}
G(\theta )=\left\{\begin{matrix} G_{0}-3.01\times (\frac{2\theta }{\theta _{-3dB}})^{2},\quad 0^{o}\leq \theta \leq \theta _{ml}/2; & \\ G_{sl}, \qquad \qquad \qquad \qquad \theta _{ml}/2< \theta \leq180^{o}; & \end{matrix}\right.
\label{eq:yi}
\end{eqnarray}
where $\theta$ denotes an arbitrary angle within the range $[0^{o},180^{o}]$, $\theta_{-3dB}$ denotes the angle of the half-power beamwidth. $\theta_{ml}$ which can be calculated as $\theta_{ml}=2.6\cdot \theta_{-3dB}$ is the main lobe width in units of degrees. The maximum antenna gain $G_{0}$ can be obtained by $G_{0}=10log(1.6162/sin(\theta_{-3dB}/2))^2$. The sidelobe gain $G_{sl}$ can be expressed as $G_{sl}=-0.4111\cdot ln(\theta_{-3dB})-10.579 \cite{fifteen}.$

Because mmWave suffer from high attenuation in non-line-of-sight (NLOS) transmissions, we use the line-of-sight (LOS) path loss model for mmWave as described in \cite{eight}. For flow \emph{i}, the received signal power at its receiver $r_i$ from its transmitter $t_i$ can be expressed as
\begin{equation}
P_{r}(t_{i},r_{i})=kP_{t}G_{t}(t_{i},r_{i})G_{r}(t_{i},r_{i})d_{t_{i}r_{i}}^{-n},
\label{eq:er}
\end{equation}
\emph{k} is a factor that is proportional to $(\frac{\lambda }{4\pi })^{2}$, where $\lambda$ is the wave length; $P_t$ is the transmission power of transmitter; $G_{t}(t_{i},r_{i})$ denotes the transmitter antenna gain from $t_i$ to $r_i$, and $G_{r}(t_{i},r_{i})$ denotes the receiver antenna gain from $t_i$ to $r_i$; $d_{t_{i}r_{i}}$ denotes the distance between  $t_i$ and $r_i$, and $n$ is the path loss exponent \cite{seven}.

We divided the interference between different flows into two cases: 1) the interference between two flows without any common node, namely, MUI; 2) the RSI after SI cancelation. The MUI caused by the transmitter $t_{l}$ of flow $l$ to the receiver $r_{i}$ of flow $i$ is defined as
\begin{equation}
P_{r}(t_{l},r_{i})=\rho kP_{t}G_{t}(t_{l},r_{i})G_{r}(t_{l},r_{i})d_{t_{l}r_{i}}^{-n},
\label{eq:san}
\end{equation}
where $\rho$ is the MUI factor related to the cross correlation of signals from different links \cite{seven}. According to \cite{sixteen}, after the cancelation of SI, the influence of RSI can be simulated by the loss of signal-to-noise ratio (SNR). So RSI can be express as $\beta_{n}N_{0}W$, where non-negative parameters $\beta_{n}$ represent the SI cancelation level of the \emph{n}th BS. Due to various factors, we assume the parameters for different BSs are different. $N_{0}$ is the onesided power spectral density of white Gaussian noise; $W$ is the channel bandwidth.

In mmWave communication, the multipath effect is reduced because of the directional transmission \cite{NY2}, and the channel can be approximated to Gaussian channel \cite{Twety}. According to Shannon's channel capacity, the data rate of flow $i$ can be recorded as
\begin{equation}
R_{i}=\eta Wlog_{2}(1+\frac{P_{r}(t_{i},r_{i})}{N_{0}W+\sum_{h}\beta _{t_{h}}N_{0}W +\sum_{l}P_{r}(t_{l},r_{i})}),
\label{eq:si}
\end{equation}
where $\eta$ is the factor that depicts the efficiency of the transceiver design, which is the range of (0,1) \cite{ten}. $h$ denotes the flow whose transmitter is the same as the receiver of flow $i$. In fact, we assume the number of $h$ is at most 1. $\beta_{t_{h}}$ is the parameter of SI cancelation level at BS $t_{h}$. $l$ denotes the flow that is transmitted simultaneously without any common node with $i$.
\subsection{Problem Formulation}
\begin{figure}
\begin{center}
\includegraphics[width=8cm]{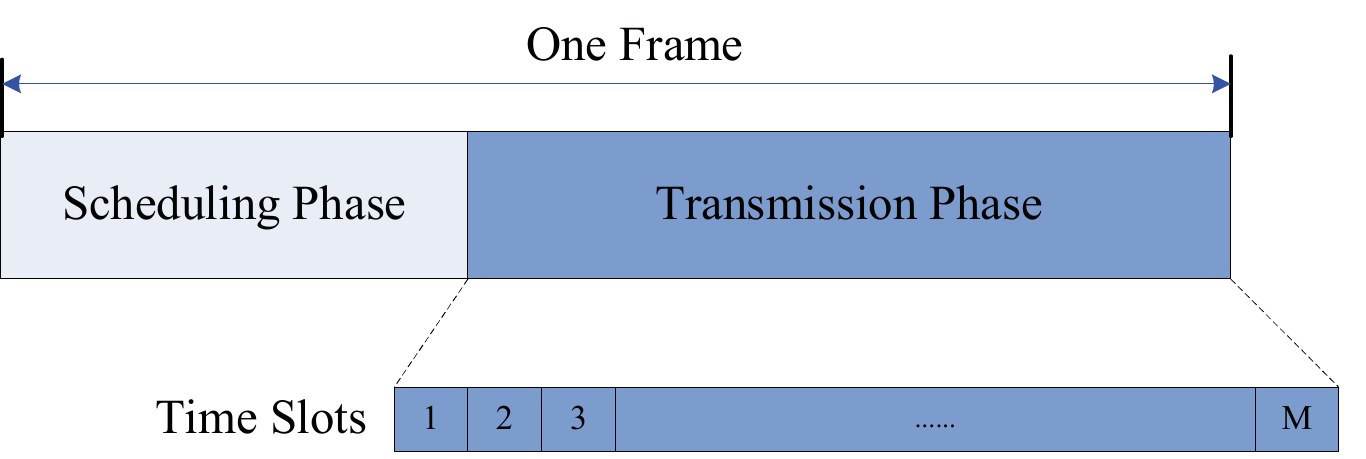}
\end{center}
\caption{The structure of one frame.}
\label{sch}
\end{figure}

 In this paper, we consider the FD concurrent scheduling problem for QoS when time slot resources are limited. The system time is divided into a series of non-overlapping frames. As shown in Fig. \ref{sch}, each frame consists of a scheduling phase, where a transmission schedule is computed by the BNC, and a transmission scheduling phase, where the BNC and BSs start concurrent transmissions following the schedule \cite{seventeen}. The transmission phase is further divided into $M$ equal TSs. We assume there are $F$ flows in the network and each flow $i$ has its QoS requirement $q_{i}$. For each flow $i$, we define a binary variable $a_{i}^{k}$ to express whether flow $i$ is scheduled in the \emph{k}th TS.
If so, $a_{i}^{k}=1$; otherwise $a_{i}^{k}=0$. Since there may be different flows to be transmitted in different TSs, we denote $R_{i}^{k}$ is the actual transmission rate of flow $i$. According to (\ref{eq:si}), $R_{i}^{k}$ can be indicated as
\begin{equation}
R_{i}^{k}=\eta Wlog_{2}(1+\frac{a_{i}^{k}P_{r}(t_{i},r_{i})}{N_{0}W+\sum_{h}a_{h}^{k}\beta _{t_{h}}N_{0}W +\sum_{l}a_{l}^{k}P_{r}(t_{l},r_{i})}.
\label{eq:wu}
\end{equation}

Then we can define the actual throughput of flow $i$ as
\begin{equation}
T_{i}=\frac{\sum_{k=1}^{M}R_{i}^{k}\Delta t}{T_{s}+M\Delta t},
\label{liu}
\end{equation}
where $T_{s}$ is the time of scheduling phase and $\Delta t$ is the time of one TS. When the actual throughput $T_{i}$ of flow $i$ is higher than its QoS requirement $q_{i}$, we deem that flow $i$ has satisfied its QoS requirement, and call the flow as a completed flow. A binary variable $A_{i}$ is used to indicate whether flow $i$ is completed. $A_{i}=1$ expresses flow $i$ is completed, while $A_{i}=0$ expresses flow $i$ is not completed.

With TDMA scheme, to satisfy the QoS requirement of flow, each flow will be allocated a greater number of time slots. In concurrent transmission, we should aim at maximizing the number of flows with their QoS requirements satisfied to better provide services in network. For this purpose, the objective function can be formulated as
\begin{equation}
max\sum_{i=1}^{F}A_{i},
\end{equation}
and the first constraint is
\begin{eqnarray}
A_{i}=\left\{\begin{matrix} 1,\qquad T_{i}\geq q_{i}; & \\ 0, \quad otherwise. & \end{matrix}\right.
\end{eqnarray}

Next, we analyze the other constraints. Firstly, we use variable $i_{n}$ to denote the flow whose transmitter or receiver
is the \emph{n}th BS $B_{n}$; so $a_{i_{n}}^{k}$ indicates whether flow $i_{n}$ is scheduled in the \emph{k}th TS, that is, whether $i_{n}$ uses $B_{n}$ in the \emph{k}th TS. In this paper, each BS is just equipped with two steerable directional antennas, so the number of flows that simultaneously using the same BS $B_{n}$ can't exceed two; this constraint can be described as
\begin{equation}
\sum_{i_{n}}a_{i_{n}}^{k}\leq 2, \forall  k,n.
\end{equation}

Then we use $i_{n}^{1}$ and $i_{n}^{2}$ indicate the two flows simultaneously using $B_{n}$; use $T(B_{n})$ and $R(B_{n})$ indicate the wireless links with $B_{n}$ as the transmitter and receiver, respectively. As assumed in section \ref{sec:System-Model} for the two antennas of a FD BS, one of them is a transmitting antenna and the other is a receiving antenna. Therefore, when two flows simultaneously use the same BS, the BS can only serve as the transmitter for one flow and as the receiver for the other, which can be expressed as
\begin{equation}
\begin{split}
i_{n}^{1}\in T(B_{n}){\rm{\& }} i_{n}^{2}\in R(B_{n}) & \\
or\quad i_{n}^{2}\in T(B_{n}){\rm{\&}} i_{n}^{1}\in R(B_{n}),
if \sum_{i_{n}}a_{i_{n}}^{k}=2.& \end{split}
\end{equation}

In short,the problem of optimal scheduling (\textbf{P1}) can be formulated as
\[max\sum_{i=1}^{F}A_{i}\]
\[s.t.\]
\[Constraints \quad (8)-(10)\]

This is a nonlinear integer programming (NLIP) problem and NP-hard \cite{Twety}. Constraint (8) indicates if the throughput of flow \emph{i} in the schedule is larger than or equal to its throughput requirement,
$A_{i} = 1$; otherwise, $A_{i} = 0$. There are many variables in constraints of our optimization problem, so the constraint is broaden. In each TS, every flow is either scheduled or unscheduled. Therefore, when the number of TSs is $M$ and the number of flow is $F$, the computational complexity using exhaustive search algorithm is $2^{MF}$, which is exponential. In FD wireless backhaul network, the number of flows may be large, and thus its complexity is unbearable for network. In the next section, we propose a FD concurrent scheduling algorithm based on coalition game (FDCG) to solve  problem \textbf{P1} with low complexity.
\section{\label{sec:coalition game algorithm}FULL-DUPLEX CONCURRENT SCHEDULING ALGORITHM BASED ON COALITIONAL GAME}
In this section, we use the idea of contention graph and the maximum independent set (MIS) \cite{ten} to solve the
situation that FD can not be directly scheduled in parallel, and to provide the basic set for coalition game algorithm which can find the maximum sum rate scheduling set. Accordingly, this algorithm can schedule concurrently flows with the maximum sum rate. Therefore, it achieves the aim of maximizing the number of completed flows and improves network throughput. Next, we introduce the relevant content of the proposed algorithm and describe it in detail.
\subsection{The Construction of Contention Graph}
\begin{figure}
\begin{center}
\includegraphics[width=8cm]{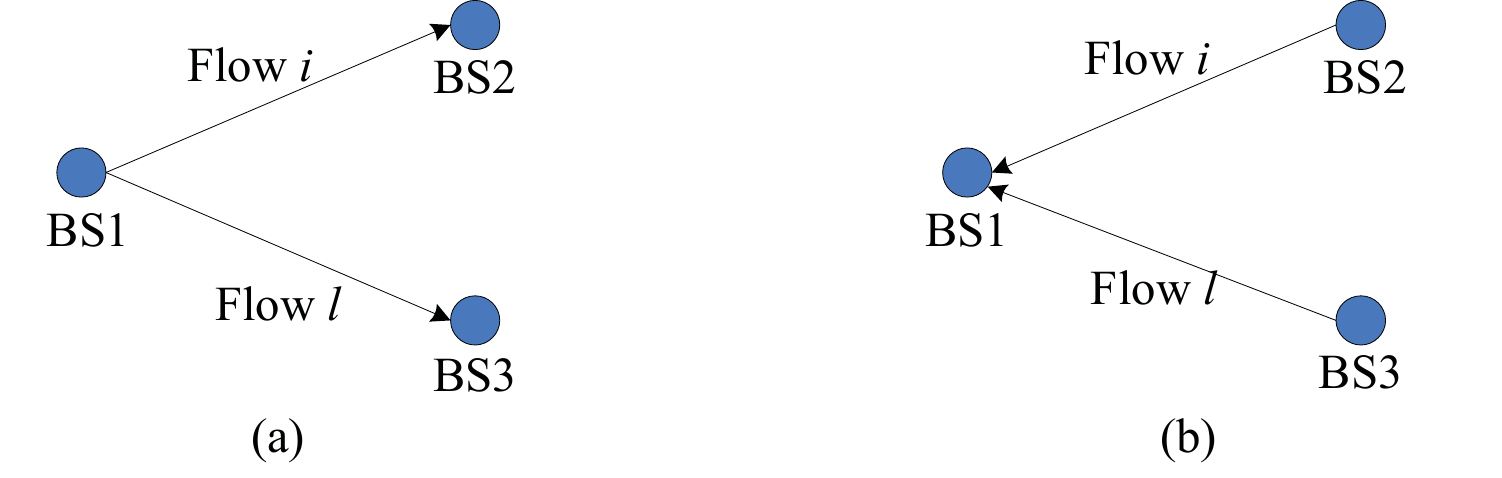}
\end{center}
\caption{The flows that can't be concurrently scheduled due to the FD assumption.}
\label{flow1}
\end{figure}

\begin{figure}
\begin{center}
\includegraphics[width=8cm]{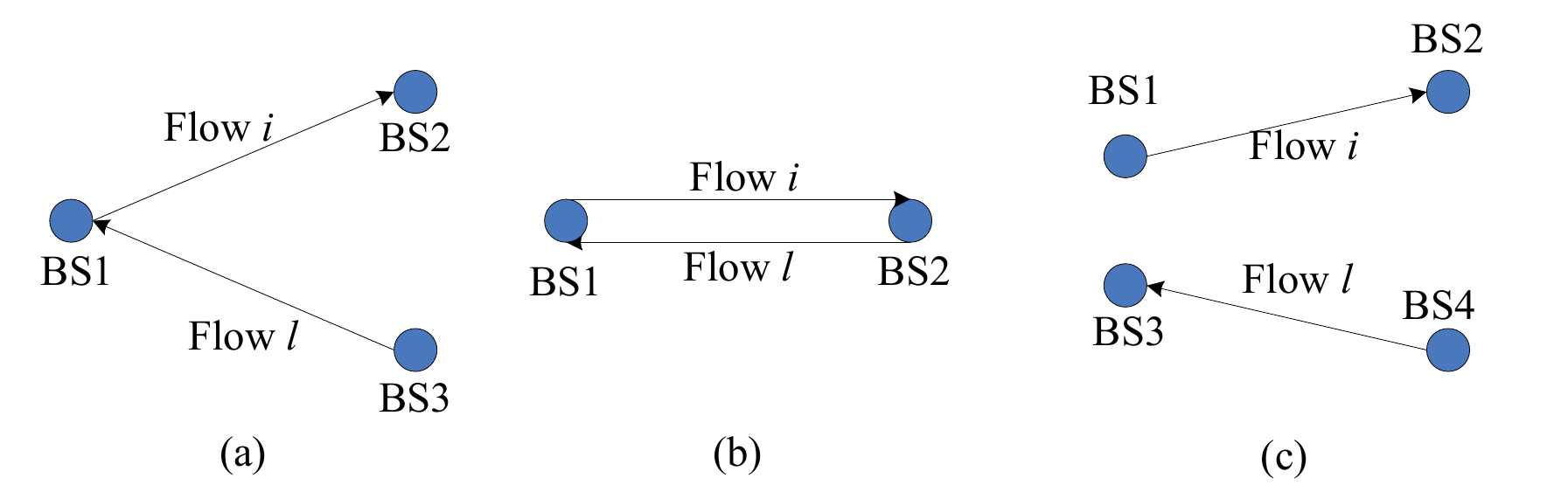}
\end{center}
\caption{The flows that allowed concurrently scheduled under the FD assumption.}
\label{flow2}
\end{figure}

In FD mmWave wireless backhaul networks, not all the flows can be concurrently scheduled. In contention graph \cite{ten}, when the two flows can't be concurrently scheduled, we define there is a contention between them. In this paper, only one case that flows can't not be concurrently scheduled is defined. According to our assumption of FD, for a FD BS with two antennas, one antenna is the transmitting antenna and the other is the receiving antenna. Therefore, the two flows using the same BS as their transmitter (or receiver) at same time can't be concurrently scheduled. This case is showed as Fig. \ref{flow1}. Fig. 4 (a) shows that two flows simultaneously use the same BS as their transmitters. Similarly, Fig. 4 (b) shows that two flows simultaneously use the same BS as their receivers. Accordingly, based on the analysis for this case, the flows that can be concurrently scheduled are divided into following three cases. 1) As shown in Fig. \ref{flow2} (a), the transmitter of flow $i$ is the receiver of flow $l$, but the receiver of flow $i$ is not the transmitter of flow $l$.
2) As shown in Fig. 5 (b), the transmitter of flow $i$ is the receiver of flow $l$, and the receiver of flow $i$ is the transmitter of flow $l$. 3) As shown in Fig. 5 (c), the transmitter of flow $i$ is not the receiver of flow $l$, and the receiver of flow $i$ is not the transmitter of flow $l$, either.

\begin{figure}
\begin{center}
\includegraphics[width=8cm]{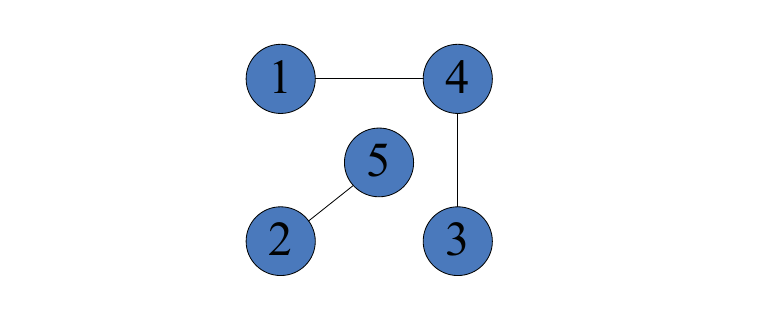}
\end{center}
\caption{The example of the contention graph.}
\label{ctu}
\end{figure}

It should be noted that noise interference, multi-user interference (MUI) and residual self-interference (RSI) are considered in this paper. Each flow has noise interference, while MUI and RSI depend on the situation. For the case in Fig. 5 (a), the interference from flow $l$ to flow $i$ is MUI, and the interference from flow $i$ to flow $l$ is RSI ;
 the interference of flow $i$ and $l$ is both RSI in Fig. 5 (b); and in Fig. 5 (c) the interference of flow $i$ and $l$ is both MUI. We need to distinguish the different interference when calculating the actual rate of the flow. Next, we construct the contention graph. In contention graph, each vertex represents a flow. If two flows can¡¯t be concurrently scheduled (i.e., there is a contention between them), an edge is inserted between the two corresponding vertices. For example, as shown in Fig. \ref{ctu},there is a contention between flow 1 and flow 4 (i.e., one of the situation in Fig. 4). In contrast, there is no contention between flow 1 and flow 5 (i.e., one of the situation in Fig. 5).
\subsection{Maximum Independent Set}
Because the coalition game adopted in this paper does not consider the flows that can not be concurrently scheduled directly under FD condition when selecting player, it is necessary to exclude the flows which can not be concurrently transmitted before selecting player. In this paper, we use the  minimum degree of maximal independent set to achieve it.
The MIS of contention graph is the set of flows that have no edge between each other on the contention graph with the maximum cardinality \cite{ten}. For example, in Fig. 6, the set of flow 1, 2, and 3, i.e., $\left \{ 1,2,3 \right \}$, is a MIS of the contention graph. Because obtaining the MIS of a basic graph is NP complete, we adopt the minimum degree greedy algorithm to approximate the MIS \cite{eighteen}. In this paper, $G(V,E)$ is used to denote the conflict graph, where $V$ denotes the set of vertices and $E$ denotes the set of edges. If there is an edge between two vertices, the two vertices are defined as neighbors. For any vertex, its neighbor vertices are represented by $N(v)$. For any vertex $v\in V$, it is used $d(v)$ is to indicate its degree. In graph theory, the degree of a vertex of a graph is the number of edges connected to the vertex \cite{ten}.

 In proposed algorithm, because the number of TSs in the transmission phase is limited, not all requested flows can be scheduled. If the flow is scheduled in all TSs of the transmission phase without satisfing its QoS requirement, it is a waste of time slot resources. To efficiently use time slot resources, this flow isn't schedule. The number of TSs that each flow spends on completing its QoS requirement can be calculated according to the
 following formula:
 \begin{equation}
 \xi _{i}=\frac{q_{i}\ast(T_{s}+M\Delta t) }{R_{i}\ast \Delta t}.
 \end{equation}
 It is important to note that $R_{i}$ is the rate of flow $i$ without interference from others, that is,
 \begin{equation}
 R_{i}=\eta Wlog_{2}(1+\frac{P_{r}(t_{i},r_{i})}{N_{0}W}).
 \end{equation}
 \begin{algorithm}
\caption{MIS Algorithm}
1: \textbf{Input}: the contention graph $G(V,E)$; the number of TSs that each flow spends to complete its QoS requirement;

2: \textbf{Initialization}: $V_{s}=\phi; V_{us}=\phi; V_{1}=V$;

3: \textbf{While} $|V_{1}|>0$ do

4:  \quad\quad    Obtain $v\in V_{1}$ such that $d(v)=\mathop {\min }\limits_{w \in V_{1}}d(w)$;

5:   \quad\quad     $V_{s}=V_{s}\cup v$;

6:    \quad\quad    $V_{1}=V_{1}-\left \{ v\cup N(v) \right \}$;

7: $V_{us}=V-V_{s}$;

8: \textbf{Return} $V_{s}$,$V_{us}$
\end{algorithm}

The pseudo-code of the MIS algorithm is presented in Algorithm 1. $V$ is the set of all flows, $V_{1}$ denotes the set of
potentially scheduled flows, $V_{s}$ denotes the MIS of flows that selected for concurrent scheduling and $V_{us}$ is
the remaining flows except for those in the MIS. A minimum degree greedy algorithm is used to obtain the MIS, as indicated by lines 3-6. In line 4, if there is more than one flow with the minimum degree, we give priority to scheduling the flows with the small number of TSs required. The computation complexity of Algorithm 1 is $O(\left | V_{1} \right |)$.

 \subsection{Full Duplex Concurrent Scheduling Algorithm Based on Coalition Game}
 The selected MIS mentioned in the previous subsection is partitioned as players and divided into two coalitions. In the coalition game, a key point is to choose which coalition to join for all flows. Therefore, it is very important to define the preference relation. Based on the preference relation, each flow can compare and order its potential coalition. For any flow $i$, the preference relation $\succ _{i}$ is defined as complete, reflexive, and transitive binary relation over the set of all coalitions that flow $i$ can possibly form. In other word, $F_{c}\succ _{i}F_{c^{'}}$ represents the flow $i$ prefers to be a member of coalition $F_{c}$ with $i\in F_{c}$ than $F_{c^{'}}$ with $i\in F_{c^{'}}$. In this paper, we define the preference inequality for all flows as follows \cite{nineteen},
\begin{equation}
F_{c}\succ _{i}F_{c^{'}}\Leftrightarrow R(F_{c})>R(F_{c^{'}}),
\end{equation}
where $ R(F_{c})$ and $R(F_{c^{'}})$ denotes the utility of coalition $F_{c}$ and $F_{c^{'}}$,respectively. The utility
is the sum transmission rates of the flows in each coalition. Suppose there are $L$ flows in the coalition $F_{c}$, $R(F_{c})$ is expressed as
\begin{equation}
R(F_{c})=\sum_{i=1}^{L}R_{i}^{k}.
\end{equation}
It can be seen from the inequality (13) that the flow tends to join coalition whose utility is higher , that is,
 to join coalition in which the sum rates of flows is greater. So we can select a coalition with the highest sum rate for
 concurrent scheduling.

 Full-duplex concurrent scheduling algorithm based on coalition game is shown as Algorithm 2, where the flows perform switch operation until the final Nash-stable partition is achieved. It has the worst case complexity of $O(\left | N \right |^{2})$. In the proposed algorithm, the line 1 is some preparation work that BNC obtains
 the BS location (Loc), the SI cancellation level ($\beta_{n}$) of each flow and the QoS requirement of each flow.
 In line 2, we calculate the number of TSs that the flow requires when there is no interference from other flows.
 The scheduling problem we study is carried out under the condition of limited time slots. In order to improve the
   availability of TSs, we first remove the flow whose TSs required is more than $M$ TSs in line 3.
  In actual scheduling, there is interference from other flows, so the rate of flow will be further reduced, and
  the number of TSs that the flow requires will be further increased. Therefore, before scheduling, we
  estimate whether each flow can satisfy the QoS requirements in the remaining TSs, and if not, remove it. Removing flows
 not only reduces the complexity of the entire scheduling, but also saves more TSs to schedule more worthy flows which will be completed in the remaining TSs.
\begin{algorithm}
\caption{Full-Duplex Concurrent Scheduling Algorithm Based on Coalition Game}
\begin{algorithmic}[1]
\STATE BNC obtain Loc,$\beta_{n}$ and $q_{i}$;
\STATE Calculate the number of the TSs $\xi _{i}$ that the flow $i$ requires to meet the QoS requirement;
\STATE Remove $D=\left \{ i|\xi _{i}>M \right \}$;
\STATE The reminding $F$ flows are sorted incrementally according to $\xi _{i}$ and obtain the pre-scheduled set \textbf{P};
\STATE Construct contention graph $G$ for all flows in the pre-scheduled set \textbf{P};
\STATE Use Algorithm 1 to select the MIS $V_{s}$ and the remaining flows set $V_{us}$;
\STATE The $N$ flows in $V_{s}$ are randomly partition two coalition $F_{c1}$ and $F_{c2}$;
\STATE Set the current partition as $F_{cur}=\left \{F_{c1}, F_{c2} \right \}, sumthroughput=0, number=0, n=0$;
\WHILE{$n<N$}
\STATE Choose a flow $i$, define its current coalition as $F_{c}$, other coalition as $F_{c^{'}}$;
\STATE $F_{c^{'}}\cup \left \{ i \right \}\rightarrow F_{c^{''}}$;
\STATE Calculate $R(F_{c})$ and $R(F_{c^{''}})$;
\IF{$max\left \{ R(F_{c}\setminus\left \{ i \right \}),R(F_{c^{''}}) \right \}> max\left \{ R(F_{c}),R(F_{c^{'}}) \right \}$}
\STATE Perform the switch operation, $F_{c}\setminus\left \{ i \right \}\rightarrow F_{c},
F_{c^{'}}\cup \left \{ i \right \}\rightarrow F_{c^{'}}$ ,obtain new partition $F_{cur}$; $n=0$;
\ELSE
\STATE $n=n+1$;
\ENDIF
\ENDWHILE until the partition reach the final Nash stability.
\STATE Note the coalition with the maximum sum rate as $F_{c_{1}}$ and another coalition $F_{c_{2}}$;
\STATE Add the remaining flows set $V_{us}$ to the coalition $F_{c_{2}}$ , become the coalition $F_{c_{2^{'}}}$;
\STATE $k=1$;
\WHILE{$k\leq M$}
\FOR{$i$($i\in F_{c_{1}}$}
\IF{$\frac{R_{i}^{k}\Delta t+R_{i}(M-k)\Delta t+\sum_{j=1}^{k-1}R_{i}^{j}\Delta t}{T_{s}+M\Delta t}<q_{i}$}
\STATE $F_{c_{1}}\setminus \left \{ i \right \}\rightarrow F_{c_{1}}$;
\ENDIF
\ENDFOR
\STATE Calculate the total throughput of all flows in the coalition $F_{c_{1}}$  at $k$ TSs as $sumthroughput$;
\STATE Calculate successively the throughput of all flows in the coalition $F_{c_{1}}$  at $k$ TSs as $th$;
\IF{ the throughput of flow $i$ $th>q_{i}\ast (T_{s}+M\Delta t) $}
\STATE Flow $i$ is completed,$F_{c_{1}}\setminus \left \{ i \right \}\rightarrow F_{c_{1}}$;
\STATE $number=number+1$; $change=1$;
\IF{$change=1$}
\FOR{$j$($j\in F_{c_{2^{'}}}$)}
\IF{flow $j$ and other flows in coalition $F_{c_{1}}$ without the situation of Fig. 4. and $R(F_{c_{1}}\cup j)> R(F_{c_{1}})$}
\STATE Update $F_{c_{1}}\cup \left \{ j \right \}\rightarrow  F_{c_{1}} $,$F_{c_{2^{'}}}\setminus \left \{ j \right \}\rightarrow  F_{c_{2^{'}}} $;
\ENDIF
\ENDFOR
\STATE $k=k+1$,$change=0$;
\ENDIF
\ENDIF
\ENDWHILE
\RETURN $sumthroughput$ and $number$;
 \end{algorithmic}
 \end{algorithm}

    In the line 4-5 , the remaining flows are sorted in ascending order according to the number of TSs that
 the flow requires and the sorted flows are pre-scheduled set \textbf{P}. The next step is to construct contention graph for all flows in the pre-scheduled set \textbf{P}. Then the MIS $V_{s}$ is selected as the player of the coalition game
 by algorithm 1. In order to further increase the number of completed flows and improve system throughput, we don't directly un-schedule the remaining flows that are not in the MIS. Instead, we put it in a set $V_{us}$ for subsequent scheduling. Then flows in MIS $V_{s}$ are randomly divided into two coalitions. Initially, the system will choose one flow $i$. In addition to the current coalition $F_{c}$, another coalition $F_{c^{'}}$ is still possible to join for the flow $i$. The BNC has complete information of network topology. Based on these information, it calculates respectively
the sum transmission rate of coalition $F_{c}$ and $F_{c^{'}}\cup \left \{ i \right \}$. In line 13, if the switch condition meets the preference relation defined in formula (13), the system performs a switch operation. In order to achieve better convergence and reduce the complexity of the algorithm, we introduce the number of continuous unsuccessful switching operations $m$. If the switch operation is complete, $m=0$; otherwise, $m=m+1$. Until $m$ is equal to the number of flows, the final Nash stability point is reached, that is, the maximum sum rate of coalition has come up.

Next, we define the higher total transmission rate of the coalition as active and schedule it concurrently.
The other coalition is inactive. In line 20, we join $V_{us}$  to the inactive coalition $F_{c_{2}}$ and
 form a new coalition $F_{c_{2^{'}}}$. Then the coalition $F_{c_{2^{'}}}$ is sorted in ascending order according to the number of TSs. Thus, we give priority to scheduling the flow with a small number of required TSs in the inactive coalition. From line 23 to 27, we estimate whether the remaining TSs can meet the QoS requirements of the flow. In the process of judgment, it is considered that there is no interference with other flows in the next TSs, that is,
 the corresponding transmission rate is $R_{i}$. Remove the flow if the QoS requirement of the flow is not satisfied
 in the remaining TSs. From line 29 to 41, the throughput of flow which has been scheduled in active coalition is calculated in turn. If the throughput of flow satisfies its QoS requirement, the flow is complete. Then it is removed to avoid repeating scheduling. $change=1$ denotes there is a completed flow. If $change=1$, the flow in the coalition $F_{c_{2^{'}}}$ are estimated successively whether or not the flow can be scheduled in the $k+1$ TS. The conditions for scheduling are that there is no contention between the added flow(s) and the scheduled flows, that is, there is no common transmitter or receiver, and that the utility of the active coalition with the joined flow(s) is higher than that of the active coalition with the unjoined flow(s). This can ensure concurrent scheduling is the maximum sum rate in each TS.
 \section{\label{sec:result analysis}PERFORMANCE EVALUATION}
 In the case of limited time, the sum rate of scheduling flows in each slot reach maximum, which can increase the number of completed flows and improve system throughput. In order to express the superiority of the algorithm more intuitively, we simulate the algorithm and analyze the result accordingly.
 \subsection{Simulation Setup}
 In the simulations, we evaluate the performance of the proposed algorithm in a 60GHz mmWave wireless backhaul network
that 10 BSs are uniformly distributed in a $100m\times 100m$ square area \cite{Twety}. Every BS has the same transmission power $P_{t}$. The transmitters and receivers of flows are randomly selected, and the QoS requirements of flows are uniformly distributed between 1Gbps and 3Gbps. The SI cancelation parameters $\beta$ for different BSs are uniformly distributed between 2 and 4. We also analyze the impact of SI cancelation level, MUI factor and different QoS requirements on the performance improvement. To be more realistic, other parameters are shown in Table I. Because we focus on the QoS of flows, according to our optimization goal, we use the number of completed flows and the system throughput as evaluation metrics. When one flow achieves its QoS requirement, it is called a completed flow. System throughput represents the throughput of all flows in the network per slot.
 \begin{table}[htbp]
	\caption{\label{c}Simulation Parameters}
\begin{center}
	\begin{tabular}{lcc}
		\toprule
		\textbf{Parameter} & \textit{\textbf{Symbol}} & \textit{\textbf{Value}}\\
		\midrule
		{Transmission power} & $P_{t}$ & 1000mW \\
		{Path loss exponent} & $n$ & 2 \\
        {Transceiver efficiency factor} & $\eta$ & 0.5 \\
        {System bandwidth}&$W$ & 1200MHz \\
        {Background noise} & $N_{0}$ & -134dbm/MHz \\
        {Slot time}& $\Delta t$ & 18us \\
        {Scheduling phase time}&$T_{s}$ & 850us \\
        {Half-power beamwidth}&$\theta _{-3dB}$ & $30^{o}$ \\
		\bottomrule
	\end{tabular}
\end{center}
\end{table}

In order to show the advantages of the proposed full-duplex concurrent scheduling algorithm based on coalition game
in mmWave wireless backhaul network, we compare it with the following three algorithms.The coalition game algorithm
proposed under FD condition is called FDCG algorithm.

\textbf{1)TDMA:} In TDMA, flows are transmitted serially. Only one stream is allowed to be scheduled in each slot during the transmission phase. we use TDMA as the baseline for evaluating performance without concurrent scheduling.

\textbf{2)STDMA:} This algorithm allows both interference and non-interference flows to be transmitted concurrently. The differences between this algorithm and the proposed algorithm are that STDMA is under half-duplex condition and doesn't choose the maximum sum rate of flows in the first TS.

\textbf{3)SFD:} This is an algorithm that removes the part of the coalition game and does not consider whether the added flows increase the sum rate in the following TSs. The flows in the MIS are scheduled directly in the first TS. When a flow is completed, the MIS can be added into the flow that only has no conflict with the scheduled flows. It is a simple FD algorithm. The proposed FDCG algorithm is compared with SFD to show the idea that coalition game is more beneficial to scheduling.

For the $M$ TSs and $F$ flows, the computational complexity of TDMA , STDMA, SFD and FDCG is $O(M)$, $O(kF)$£¬$O(MF)$ and $O(MF)$, respectively, where $k$ is a constant coefficient.
 \subsection{Performance Evaluation}
In order to make the simulation data more persuasive, each simulation performs 10 times.
The topology of the network, the location of the base station and the QoS requirements of all flows vary with
 each simulation in this paper. From Fig. 7 to Fig. 14, MUI factor is 1.
 \begin{figure}
\begin{center}
\includegraphics[width=8cm]{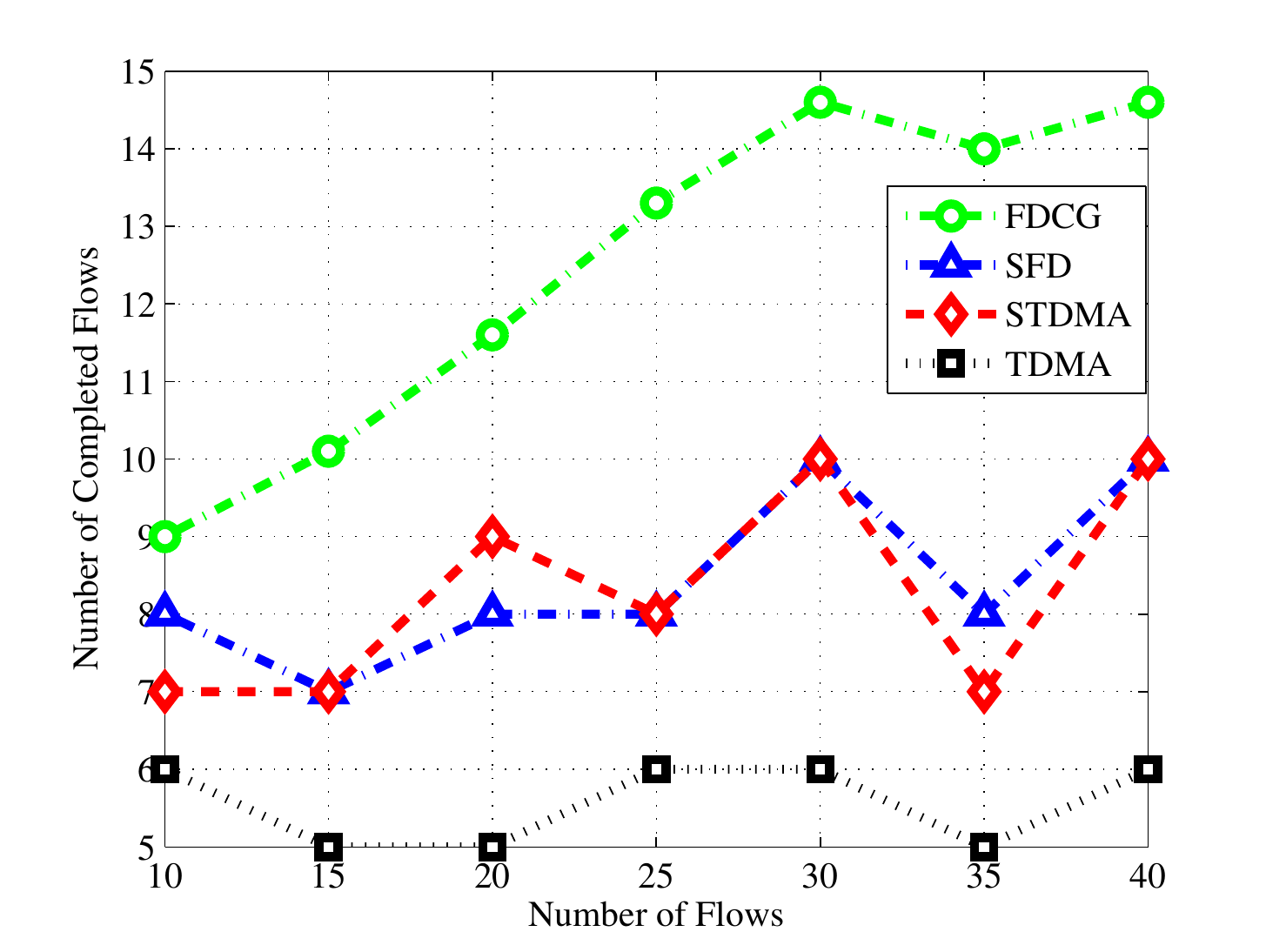}
\end{center}
\caption{Number of completed flows comparison of four scheduling algorithm with different number of time slots.}
\label{re1}
\end{figure}
\begin{figure}
\begin{center}
\includegraphics[width=8cm]{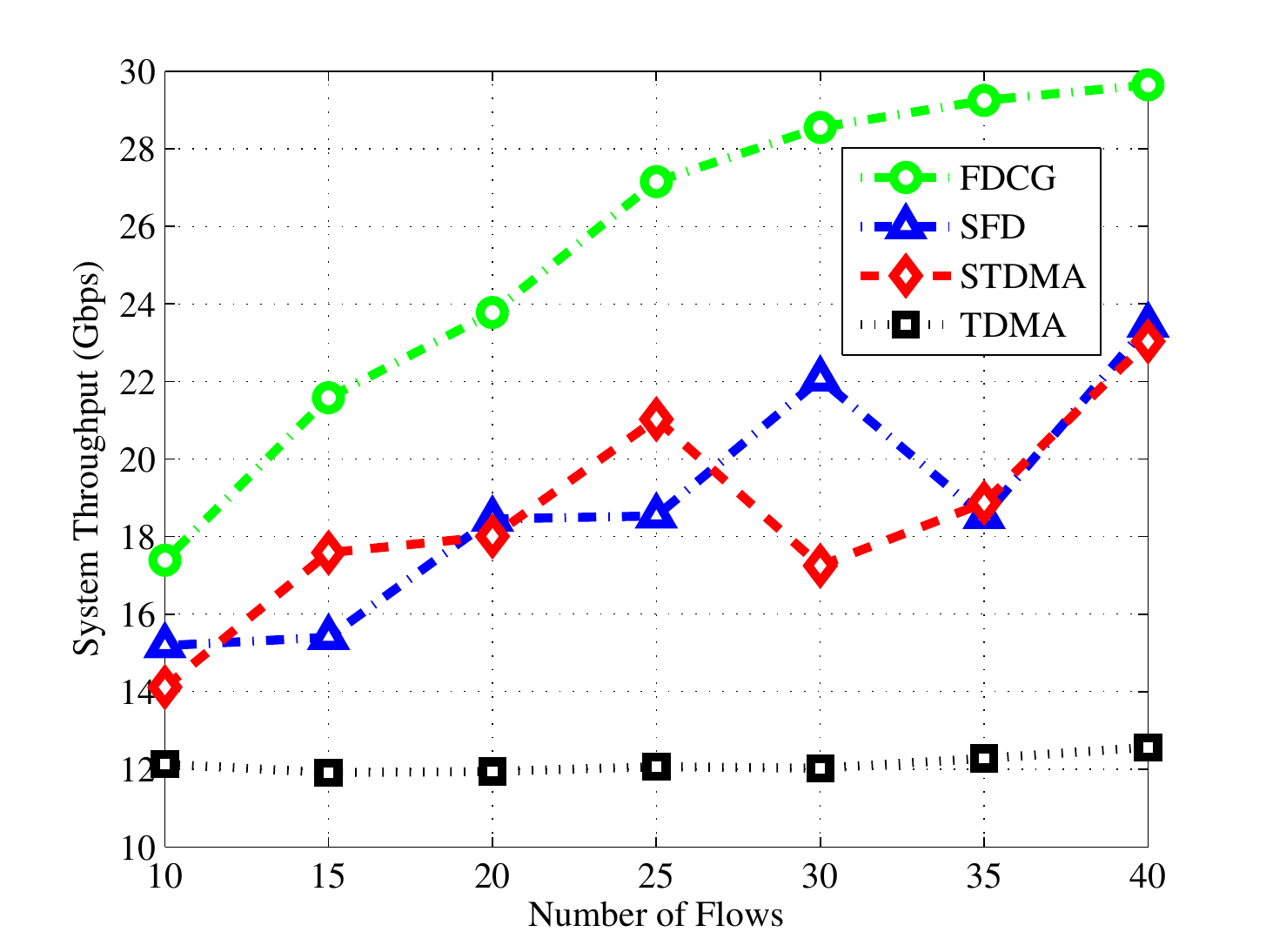}
\end{center}
\caption{System throughput comparison of four scheduling algorithm with different number of time slots.}
\label{re2}
\end{figure}
 In Fig. \ref{re1}, we set the number of time slots to be 1000. Then, we plot the number of completed flows comparison of
 four scheduling algorithm varying the number of flows from 10 to 40. From the figure, we can observe the proposed
 FDCG algorithm is better than the other three. For TDMA, when the number of flows increases and the
 number of TSs remains constant, the number of completed flows remains essentially the same. Because only one
 flow per slot can be transmitted in TDMA. The number of completed flows in STDMA algorithm fluctuates
 greatly because the QoS requirements of the flows and the position information are random in every simulation.
 And the STDMA and SFD have little change with the increase of the number of flows. For the proposed FDCG algorithm
 in this paper, with the increase of the number of flows, the number of completed flows is also increasing. When the number of flows is over 30, the number of completed flow tend towards stability because of the limited time slots resources. It shows that the proposed algorithm based on coalition game is more suitable for the current situation in which the amount of  traffic data and the number of users increase.
  \begin{figure}
\begin{center}
\includegraphics[width=8cm]{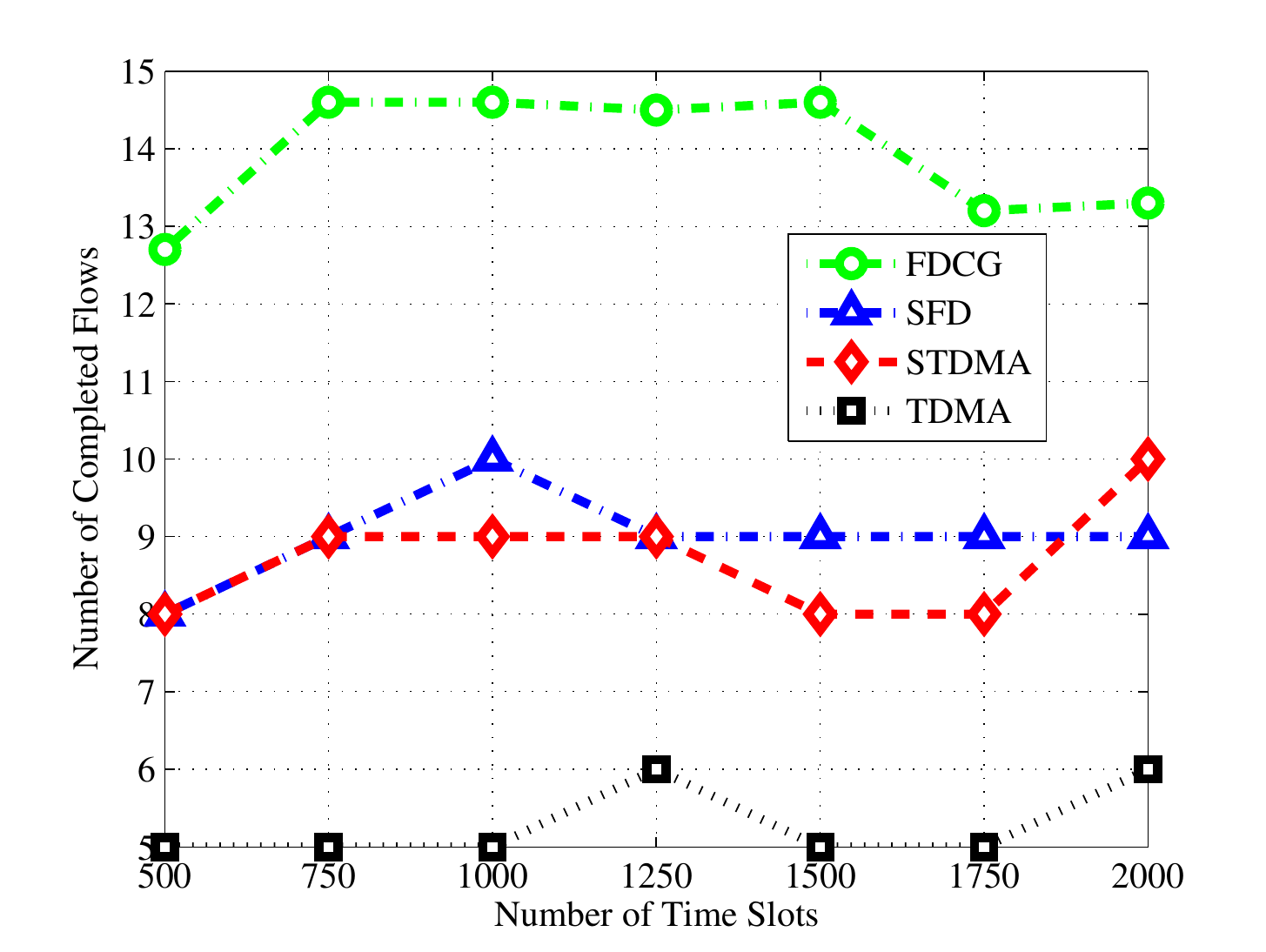}
\end{center}
\caption{Number of completed flows comparison of four scheduling algorithm with different number of flows.}
\label{re3}
\end{figure}
\begin{figure}
\begin{center}
\includegraphics[width=8cm]{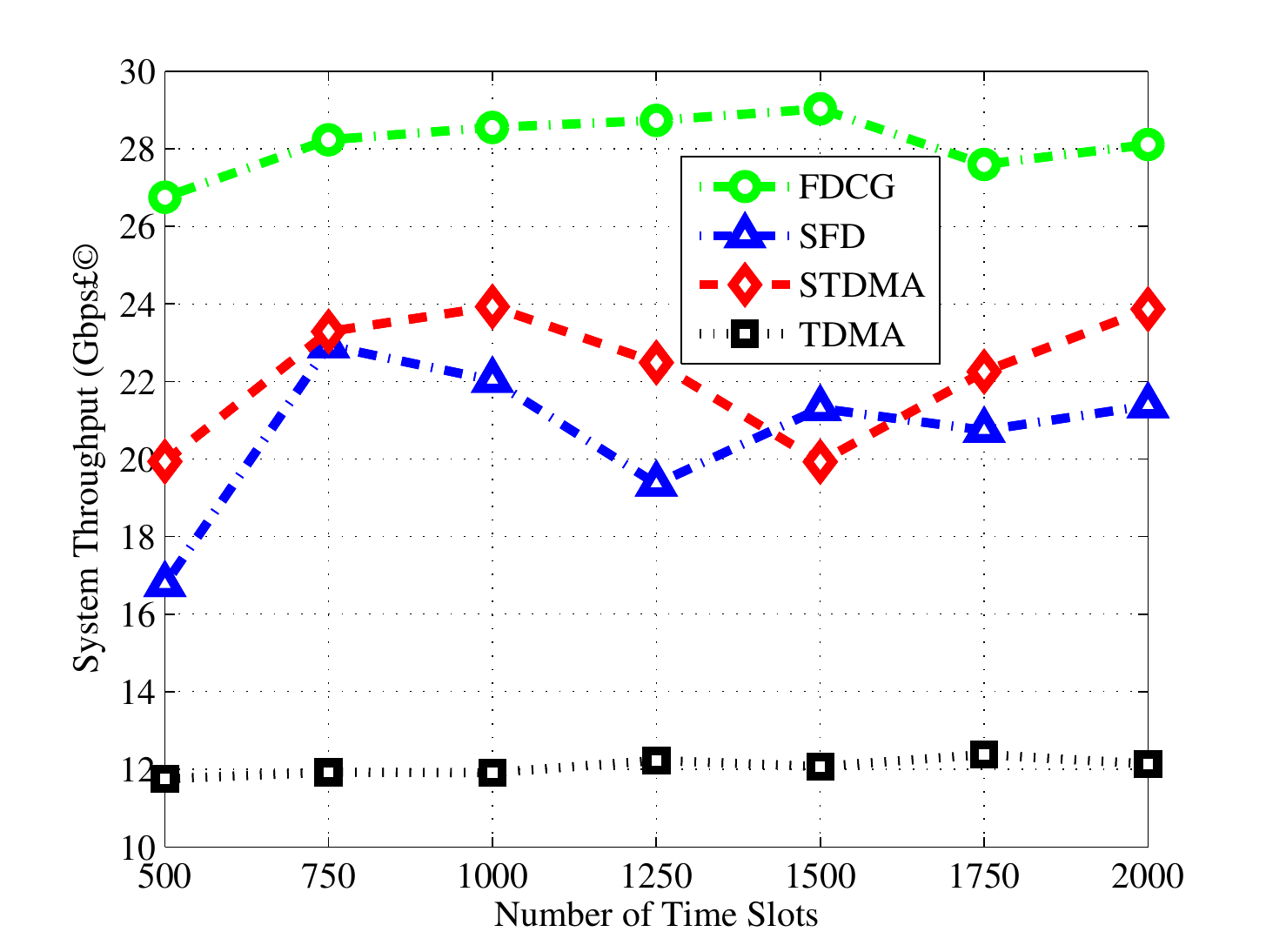}
\end{center}
\caption{System throughput comparison of four scheduling algorithm with different number of flows.}
\label{re4}
\end{figure}
 In Fig. \ref{re2}, we set the number of time slots to be 1000. Then, we plot the system throughput comparison of
 four scheduling algorithm varying the number of flows from 10 to 40. The system throughput by TDMA remains around
  12Gbps, mainly because the algorithm schedules only one flow per slot and the system capacity is limited.
  Therefore, no matter how the number of flows changes, the system throughput will not increase too much.
  The throughput of the proposed FDCG algorithm is about 30Gbps, and the lower throughput is 17.5Gbps.
  When the flow number is 30, the throughput of FDCG is higher than that of SFD, STDMA and TDMA about 29.31\%, 65.45\%
  and 137.24\%, respectively.

In Fig. 7 and Fig. 8, the proposed FDCG algorithm is superior to the other three algorithms in both the number of
completed flows and system throughput. FDCG is better than SFD because FDCG has the maximum sum rate
in each slot. Moreover, we also give the priority to scheduling the flows with small number of required TSs, which will save more time slots to satisfy the subsequent flow transmission. When a flow is completed, the interference of other flows will also reduce and the rate of each flow will increase. Therefore, the flows  will spend less time slots on satisfying their QoS requirements , which is a virtuous cycle. Due to the direct scheduling of the MIS by SFD, many flows are scheduled concurrently , which leads to the increase of the total interference of each flow and the decrease of the transmission rate of each flow. For this reason, more time slots are needed, resulting in the long-term occupation of time slot resources. For STDMA, the algorithm is based on half-duplex condition, and concurrent scheduling flows are not the maximum sum rate of flows. Even if the rate is considered in the subsequent joined flow, the scheduled flows will not become the maximum sum rate. As a result, the number of completed flows and the system throughput of STDMA are not as good as that of FDCG algorithm.

In Fig. \ref{re3}, we set the number of flows to be 30.Then, we plot the number of completed flows comparison of four
scheduling algorithms varying the number of time slots from 500 to 2000. From the figure, the FDCG algorithm achieves
significantly better performance than other algorithms. When the number of time slots is 1500, the number of completed
flows of the FDCG algorithm is higher than that of SFD£¬STDMA and TDMA about 62.22\%, 82.50\% and 192.00\%.
However, with the increasing number of slots up to 1500, the number of completed flows of FDCG decreases. Because
the QoS requirement of each flow is converted to the number of bits in the simulation and the number
of bits transformed is proportional to the number of time slots. That is to say, the larger the number of time slots is the greater the number of bits per flow. So flows require more time slots to satisfy their QoS requirements, which may lead to many later scheduled flows do not satisfy their QoS requirements. This results in a reduction of the number of completed flows.

In Fig. \ref{re4}, we set the number of flows to be 30. Then, we plot system throughput comparison of four
scheduling algorithms varying the number of time slots from 500 to 2000. The system throughput of FDCG algorithm is obviously better than that of the other three algorithms. With the increasing number of time slots, the system throughput of TDMA algorithm does not increase significantly,but FDCG, SFD and STDMA have an obvious increasing. However, with the increasing number of time slots up to 1500, the system throughput of FDCG decreases. It can be seen that with the increasing number of slots from the formula (6), its effect on improving system throughput is negative.

\begin{figure}
\begin{center}
\includegraphics[width=8cm]{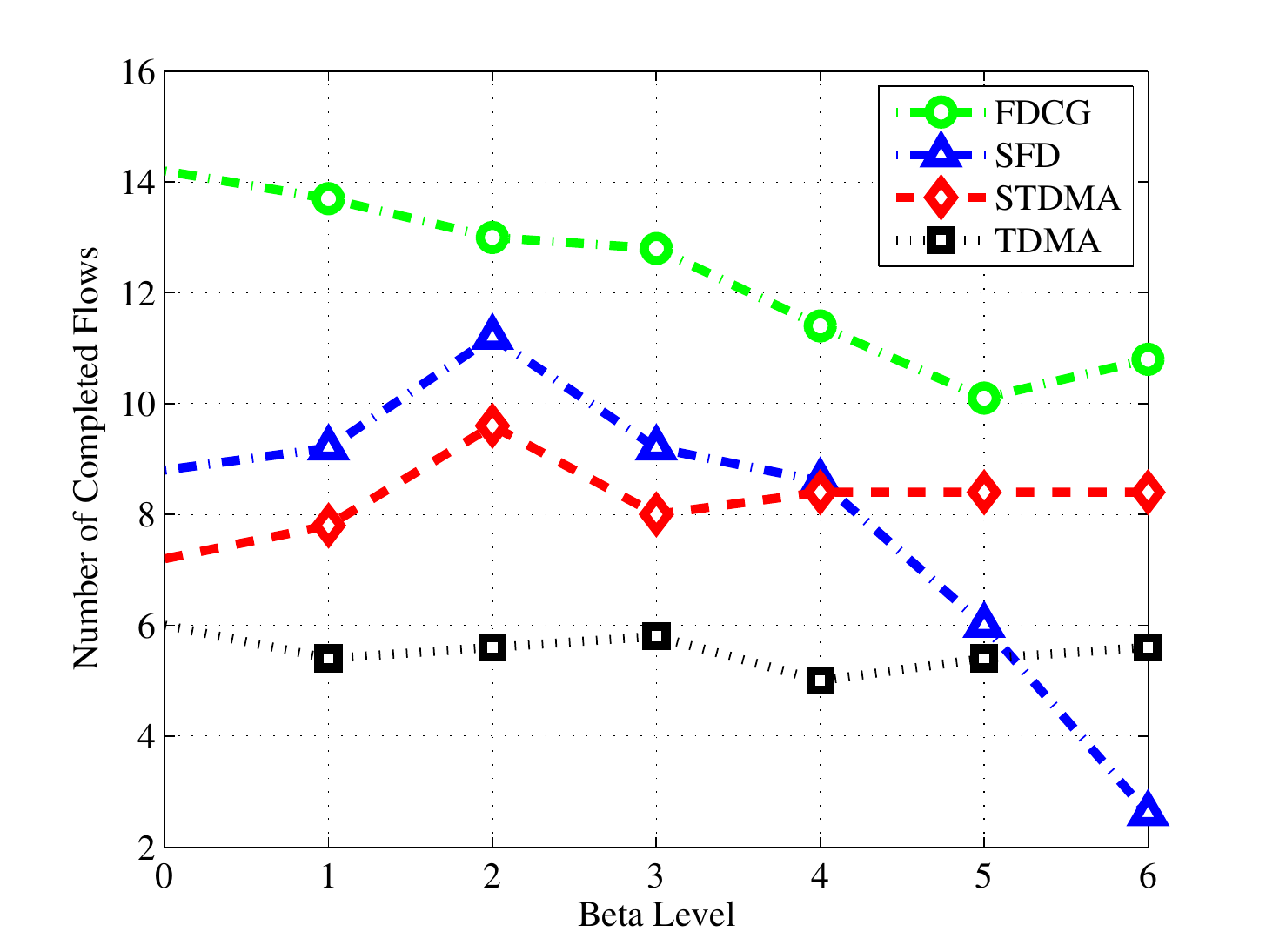}
\end{center}
\caption{The number of completed flows comparison of four scheduling algorithm with different SI cancelation level.}
\label{betaflow}
\end{figure}
\begin{figure}
\begin{center}
\includegraphics[width=8cm]{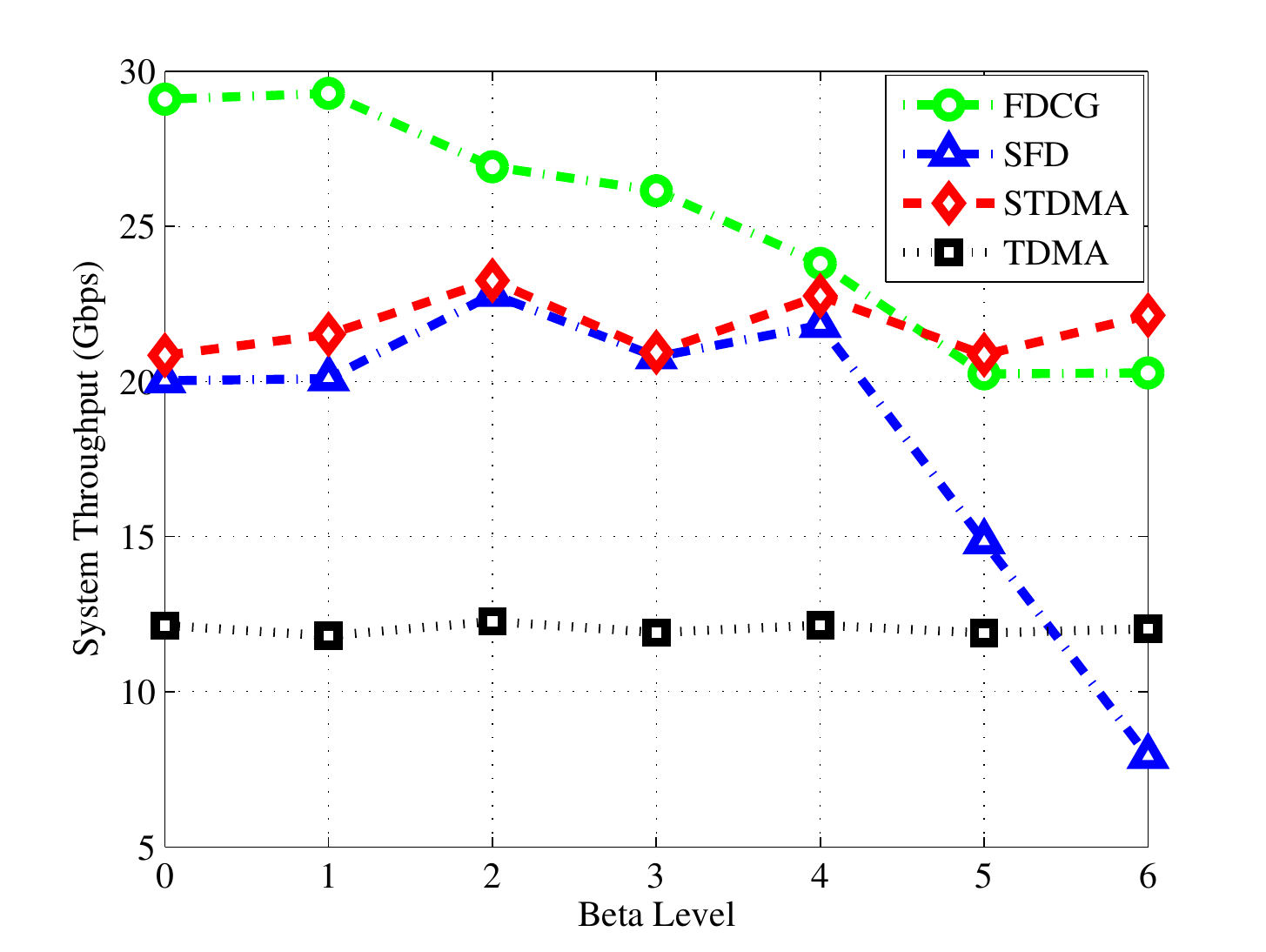}
\end{center}
\caption{System throughput comparison of four scheduling algorithm with different SI cancelation level.}
\label{betathroughput}
\end{figure}
From Fig. 9 and Fig. 10, it can be seen that the three algorithms, FDCG, SFD and STDMA, are better in terms
of the number of completed flows and the system throughput in the time slot range of 750-1500. The number of completed flows for SFD is higher than that of STDMA, but its  system throughput is lower than that of STDMA . The two algorithms schedule the flows in different ways. SFD algorithm give priority to scheduling flows with the small number of required TSs in following TSs. The completed flows are just the flows of low QoS requirements, resulting in a small increase in system throughput. Although the number of completed flows is high, the system throughput is not necessarily high. At the same time, it can be seen that the FD has twice as much spectrum as the half-duplex, but if the scheduling algorithm is not good enough, it will be better or worse than the half-duplex scheduling algorithm. This also reflects the importance of scheduling algorithms.

In Fig. \ref{betaflow} and Fig. \ref{betathroughput}, for the proposed FDCG and SFD, the SI cancellation level $\beta$ has an obvious effect on the performance. We simulate the performance under different magnitudes of $\beta$. For example, when x=3, $\beta$ is uniformly distributed in $2\times 10^{3}-4\times 10^{3}$. In this case, the total number of flows is 40 and $M=1000$. We can find the performance of SFD deteriorates rapidly after $\beta$ reaches $10^{4}$ magnitude, but the proposed FDCG gradually deteriorates. When $\beta$ reaches $10^{5}$ magnitude, the proposed FDCG has the similar performance as the STDMA. This also indicates the advantages of full duplex will not show up until SI cancelation level is below a certain value.

\begin{figure}
\begin{center}
\includegraphics[width=8cm]{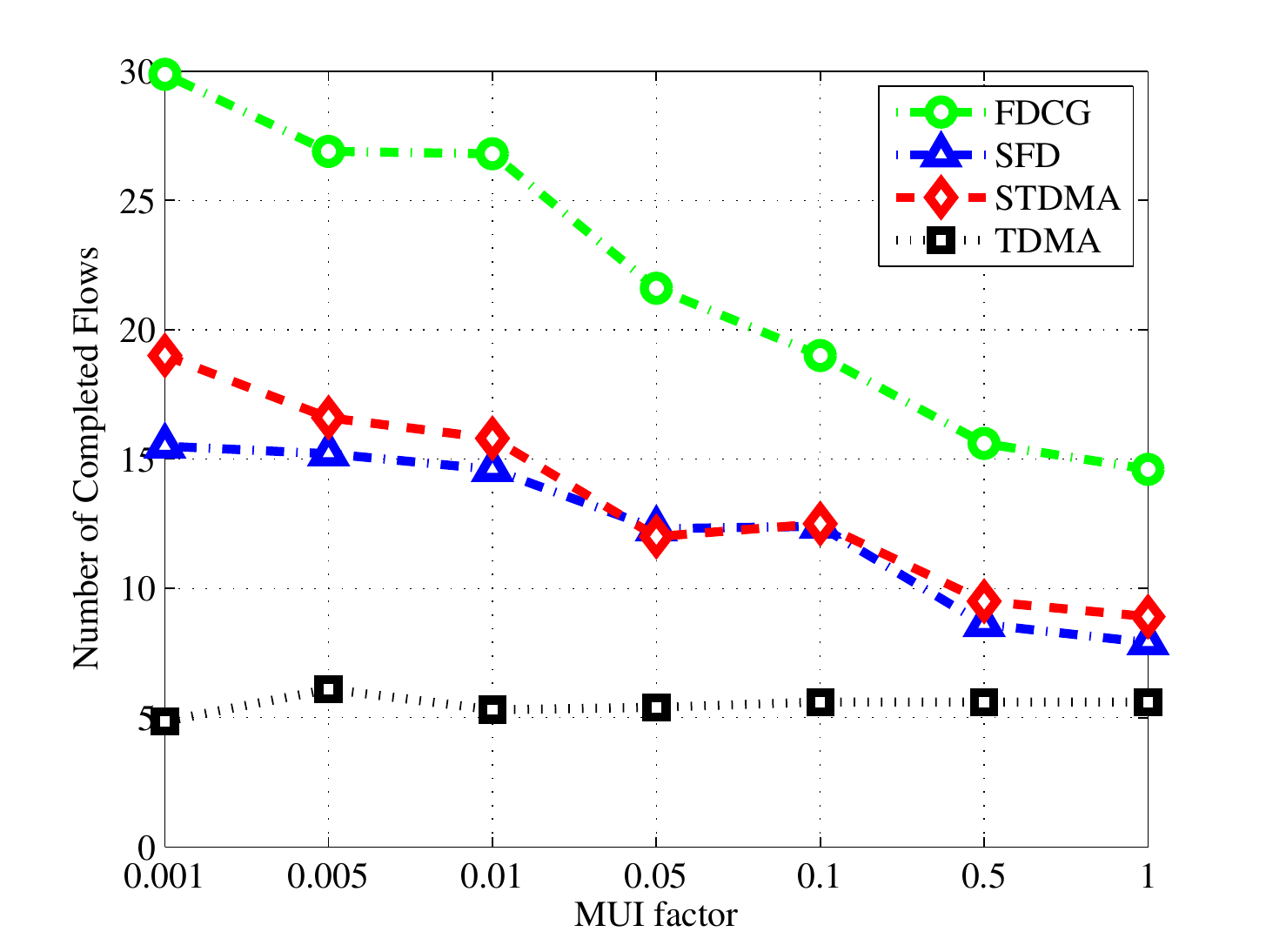}
\end{center}
\caption{The number of completed flows comparison of four scheduling algorithm with different MUI factor.}
\label{MUIflow}
\end{figure}

\begin{figure}
\begin{center}
\includegraphics[width=8cm]{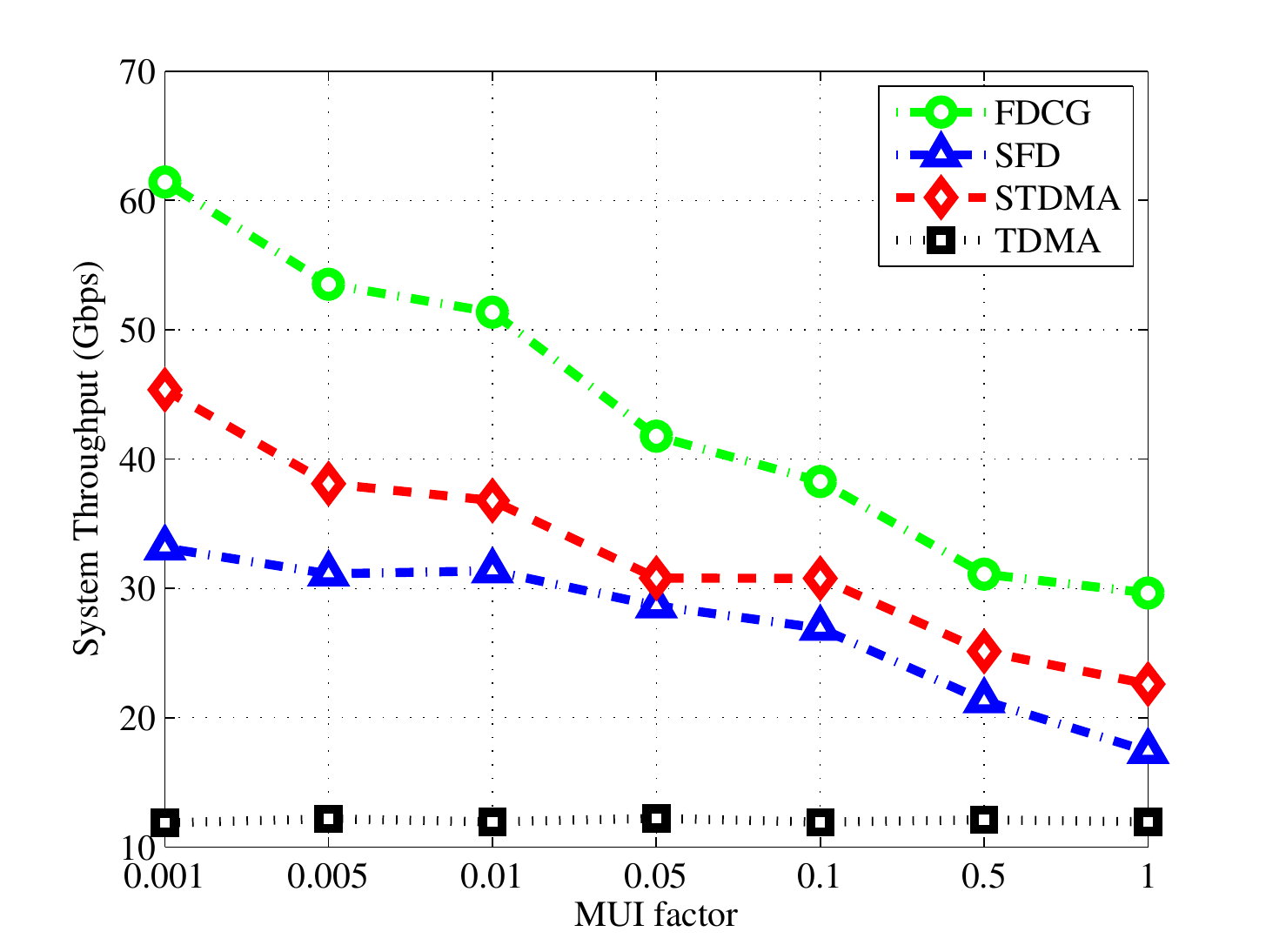}
\end{center}
\caption{System throughput comparison of four scheduling algorithm with different MUI factor.}
\label{MUIthroughput}
\end{figure}
In Fig. \ref{MUIflow} and Fig. \ref{MUIthroughput}, for the proposed FDCG , SFD and STDMA, the MUI factor $\rho$ has an obvious effect on the performance. In this case, the total number of flows is 40 and $M=1000$. With the increasing value of MUI factor, the number of completed flows and system throughput in three algorithms all decrease in the same trend. But the performance of the proposed FDCG algorithm is higher. When $\rho=0.05$, the number of completed flows of the proposed FDCG algorithm is higher than that of STDMA about 61.53\%.

\begin{figure}
\begin{center}
\includegraphics[width=8cm]{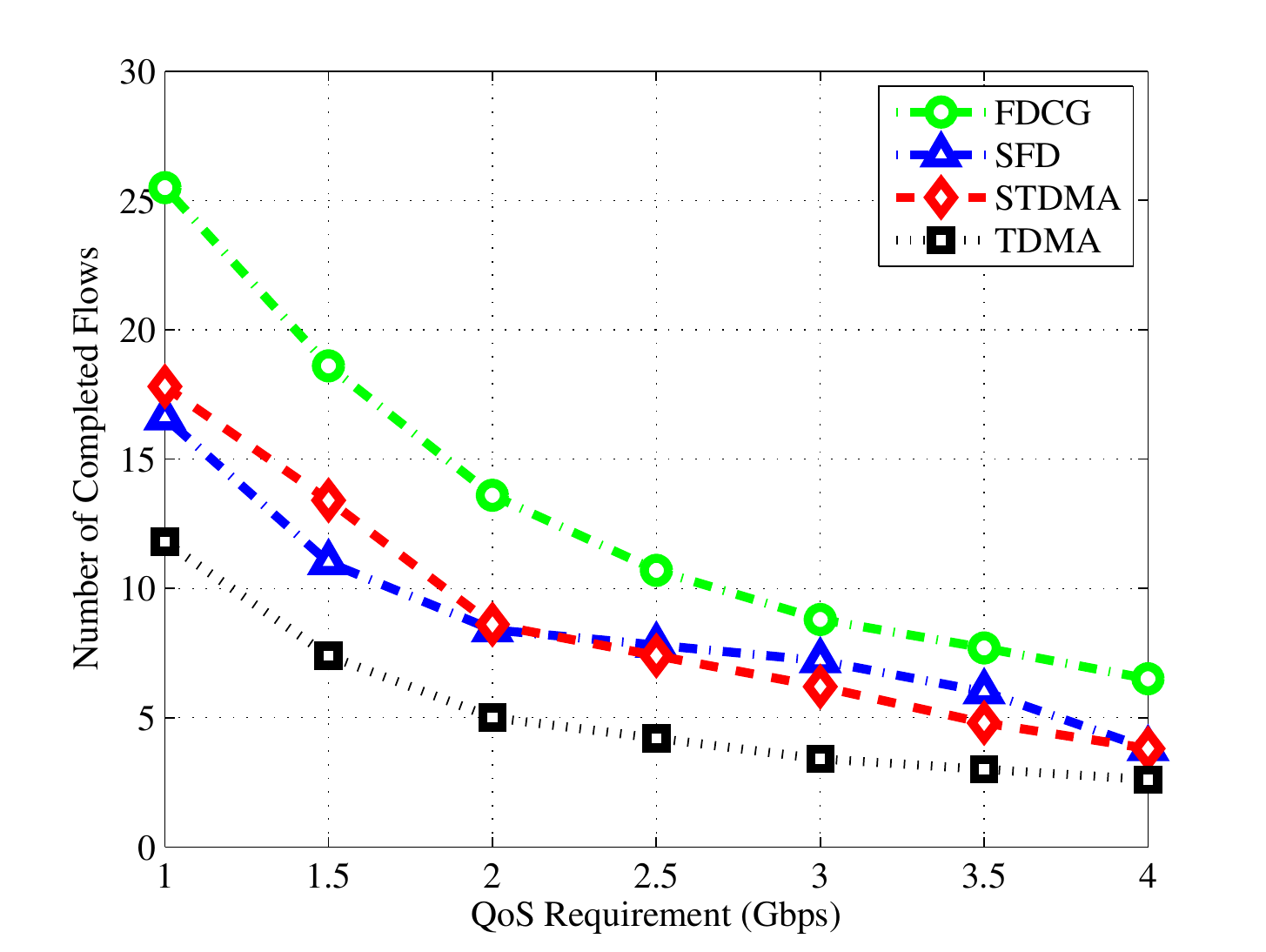}
\end{center}
\caption{The number of completed flows comparison of four scheduling algorithm with different QoS requirement.}
\label{qosflow}
\end{figure}
\begin{figure}
\begin{center}
\includegraphics[width=8cm]{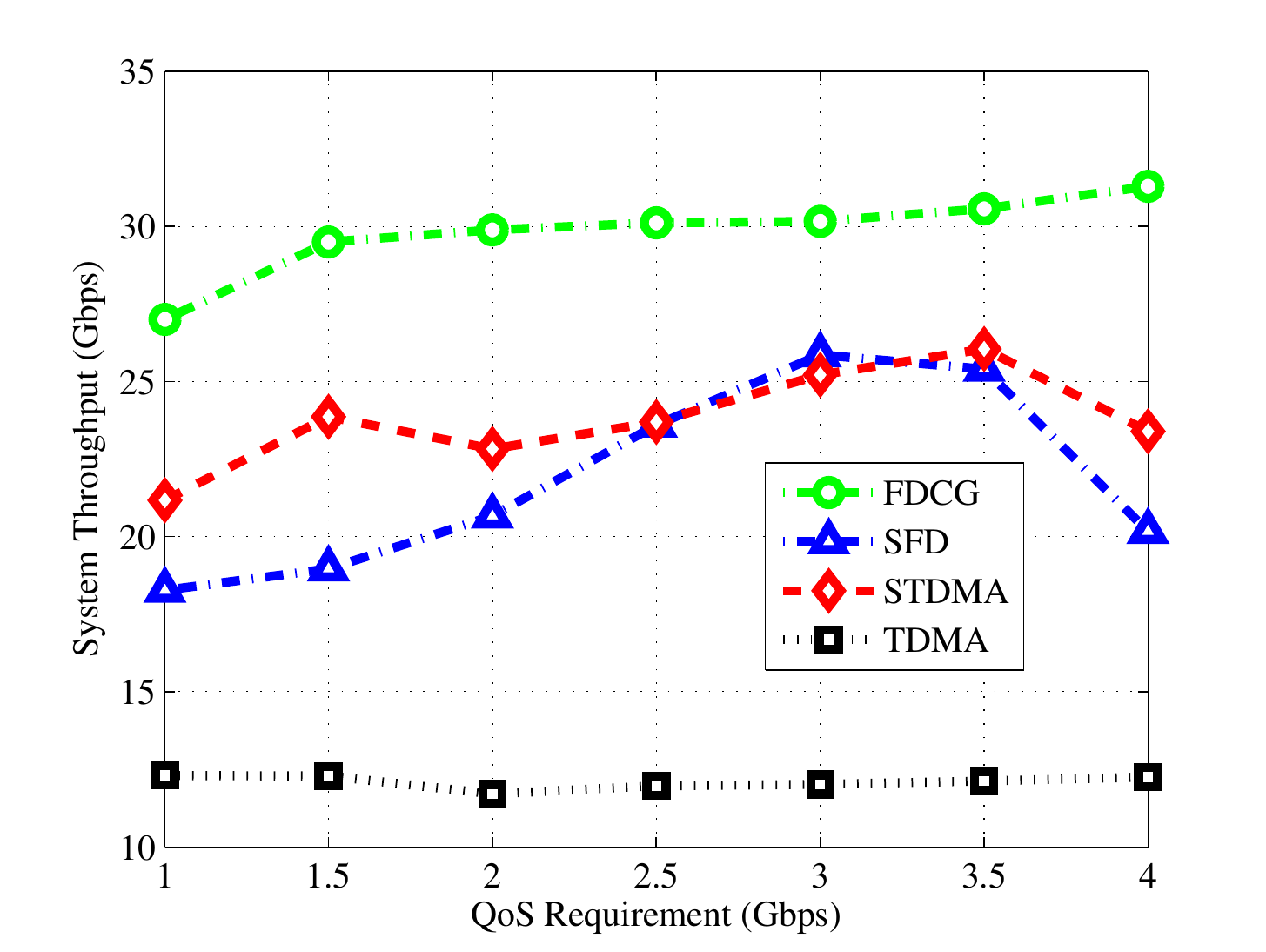}
\end{center}
\caption{System throughput comparison of four scheduling algorithm with different QoS requirement.}
\label{qosthroughput}
\end{figure}
 We set the same Qos requirement for the 40 flows and $M=1000$, and observe the performance changes of the four algorithms with the increasing QOS requirement. In Fig. \ref{qosflow}, with the increase of QoS requirement, the number of completed flows for the four algorithms decreases. When QoS requirement is 4 Gbps, the number of completed flows for the proposed FDCG is still higher than that of STDMA , SFD and TDMA. In Fig. \ref{qosthroughput}, with the increasing  QoS requirements, the system throughput of the proposed FDCG increases by a small margin and keep around 30 Gbps. For the proposed FDCG algorithm, the increasing  QoS requirements of flows has little effect on system throughput. But if the QoS requirements of every request flow is high, the number of completed flows is small.

\section{\label{sec:conclution}CONCLUSION}
In this paper, we investigate the problem of concurrent transmission scheduling in mmWave wireless backhaul network. We
propose a full-duplex concurrent scheduling algorithm based on coalition game to solve the problem.
The proposed algorithm is to use the idea of coalition game to select the concurrent scheduling set which has the maximum sum rate. The simulation results show that the proposed algorithm is superior to the other three algorithms in terms of the number of completed flows and the system throughput. In general, the proposed algorithm can increase system throughput and improve resource utilization efficiency.

\bibliographystyle{IEEEtran}

\bibliographystyle{IEEEtran}

\end{document}